  \documentstyle[epsf,aps,multicol]{revtex}
\begin{document}
\date{\today}
%
%
%
\let\makelabel=\descriptionlabel

\draft
%
\title{Modes of counterion density-fluctuations and 
 counterion-mediated   
 attractions between like-charged fluid membranes}
\author{Bae-Yeun Ha}
\address{Department of Physics, Simon Fraser University, Burnaby, B.C., V5A 1S6, Canada } 

\maketitle

\begin{abstract}
Counterion-mediated attractions between like-charged fluid
 membranes are long-ranged and non-pairwise additive at high
 temperatures.  At zero temperature, however, they are 
pairwise additive and decay exponentially with the membrane separation.
Here we show that the nature of these attractions is determined by
the  dominant modes of fluctuations in the density of counterions.
 While the non-pairwise additive interactions arise from long-wavelength
 fluctuations and vanish at zero temperature, the short-ranged
 pairwise additive interactions arise from short-wavelength fluctuations
 and are stronger at low temperatures.
\end{abstract}
\begin{multicols}{2}
 
\narrowtext
 Counterion-mediated attractions play a significant role
 in many physical and biological
phenomena~\cite{ha.2rod,ha.bundle,jensen,joanny,bloomfield.curr,janmey,kardar,pincus,safran,thirum}. These attractions are responsible for fitting DNA inside a small
 biological
 container such as a viral capsid or a nuclear
 envelope~\cite{bloomfield.curr,klimenko} and can also be crucial in
 promoting adhesion and fusion of biological
 membranes~\cite{jacob.science}.  Accordingly, significant effort has
 been expended in developing a practical way of investigating the nature of
 these counterion-mediated attractions.   Besides an integral equation method~\cite{marcelja}, there have emerged two distinct 
approaches.  The first approach~\cite{manning,oosawa}, based on a charge fluctuation 
picture, suggests that these attractions are mediated by correlated
fluctuations of ion clouds of counterions.  This approach is 
consistent with our conception of counterions as fluctuating objects 
and thus merits significant consideration~\cite{ha.2rod,ha.bundle,joanny,kardar,pincus}.  In the 
second approach, based on a zero temperature 
picture~\cite{bloomfield,levin.preprint,cruz,leikin,shklovskii}, the appearance of attractions between 
like-charged molecules is attributed to the strong charge correlations that drive 
the systems, together with counterions, into an ionic crystal.  At 
first glance these two approaches appear to be contradictory to one 
another, 
but there is an evidence that they can, in fact,  be 
complimentary~\cite{ha.correlation,ha.pre,ha.reply}.  Despite this, there 
still remain fundamental discrepancies 
between the two that have yet to be resolved.   For the case of two 
planar surfaces a distance $h$ apart, the zero-temperature 
picture leads to an attractive force that decays expenentially with $h$~\cite{bloomfield}.
 In the charge-fluctuation approach, however, the attractive force 
decays algebraically as
 $h^{-3}$~\cite{kardar,pincus}, as long as $h$ is sufficiently large.  Most recently, the apparent discrepancy
 between these two approaches
 has been examined~\cite{lau.plasmon} by Lau et al.  Lau et al. have shown
 that
 the zero-temperature quantum fluctuations give rise to a long-ranged
 attraction which varies as $h^{-7/2}$.  Their theory, however, does
 not explain the crossover between the long-ranged interactions driven
 by thermal fluctuations and the exponentially-decaying interactions
 expected at low temperature.  At finite temperatures,
 the long-ranged interactions supported by the charge-fluctuation
approach should exist and constitute a dominant contribution to the
 plate-plate interactions.  Furthermore,  when applied to many rod
 systems~\cite{ha.bundle,ha.pre,podgornik}, the charge-fluctuation
 approach suggests that interactions between rods are {\it not} pairwise
 additive while the exponentially decaying interactions between
 plates as implied by the zero-temperature approaches are pairwise additive.

In this Letter, we present a simple theoretical approach to
 counterion-mediated interactions between fluid membranes in order to
bridge the gap between the two existing theories.  In particular, we show
 that the nature of these attractions is controlled  by the dominant modes of 
 fluctuations in the density of counterions.  At high temperature, the membrane
 interactions are dominated by {\it long}-wavelength fluctuations in
 the counterion density, {\i.e.}, fluctuations at large length scales.
 The resulting {\it long}-wavelength interactions are shown to be
 {\it long}-ranged and {\it non}-pairwise additive.  As the temperature decreases, the 
 high-temperature behavior of the membrane interactions crosses over to the behavior 
 expected at low temperatures.  At low
temperatures,  the membrane interactions are mainly determined by
  the {\it short}-wavelength charge fluctuations.  We find
 that the resulting interactions decay
 {\it exponentially} with the separation between the membranes {\it and} are
 approximately pairwise additive.  Finally, we obtain a phase diagram 
 to depict the two 
 distinctive regimes characterized by  the corresponding  dominant modes, and 
 the crossover boundaries between the two.  

The system we consider here consists of negatively charged 
parallel membranes surrounded by neutralizing counterions of 
opposite charge.   
In the following derivation, we assume that counterions are localized 
in   
the plane of the membrane.  Thus our approach presented here is relevant 
to the {\it strongly} charged case.   More precisely, the separation between 
plates   
is much larger than the Gouy-Chapman length, a typical length scale within 
which counterions are confined.  The 
main purpose of the present work is to study the crossover 
from the high-temperature 
results for the membrane attractions to 
the behavior expected at low temperatures.  Since this 
crossover occurs at low temperaures or at high densities of counterions, 
the assumption of localized counterions is reasonable.

The charge distribution on a layer $j$ is described by the 
local surface charge density, $\hat \sigma_{j}({\bf r}_{\bot})= - 
e \sigma_{0} +e m Z$, where $e$ is the electronic charge, 
$\sigma_{0}$ is the average counterion number density, $m=0,1,2,3, 
$ etc., is the number of counterions per unit area at ${\bf 
r}_{\bot} \equiv (x,y) $, and $Z$ is the 
counterion valency.   
In order to calculate the free energy and the corresponding charge correlations 
(cf. Eq.~\ref{G(R)}), we use two-dimensional Debye-H\"uckel (DH) theory.  
This approximation, 
however, fails to capture the strong charge-correlations at low 
temperatures.   This defect in the DH theory has been corrected in an approximate way by 
taking into account short-ranged charge correlations over the size of 
ions~\cite{ha.correlation,ha.pre,ha.reply}.  We thus implement DH theory with  the counterion size via the two-dimensional 
form factor $g({\bf r}_{\bot}€,{\bf r}_{\bot}^{\prime})=\Theta(|{\bf 
r}_{\bot}€-{\bf r}_{\bot}^{\prime} |-D)/\pi D^{2}$ where $D$ is the diameter of 
the counterions.  We find that the 
charge-fluctuation contribution to the free energy per area is  
\begin{equation}
\label{free}
{\Delta  F_{N}€ \over k_{B}€T}={1 \over 2}  \int {d{\bf k}_\bot \over (2 \pi)^{2}€} 
\left\{ \log \Bigl[ \det Q ({\bf k}_{\bot}) \Bigr] -N { 
g({\bf k}_{\bot}€) \over \lambda 
k_{\bot}€} \right\}
,\end{equation}
where  $ \lambda^{-1}€=2 \pi Z  \ell_{B} \sigma_{0}$ is the 
inverse Gouy-Chapman 
length and $\ell_{B}=e^{2}/\epsilon k_{B}€T$ is the Bjerrum length, 
{\i.e}, the length scale at which the electrostatic energy between two 
charges is comparable to the thermal energy.  The matrix 
$Q ({\bf k}_{\bot}€)$ is defined by the matrix elements
\begin{equation}
Q_{ij}€ ({\bf k}_{\bot}) =\delta_{ij} +{ g_{ij}({\bf k}_{\bot}€) \over \lambda 
k_{\bot}€} {\rm e}^{-{ k}_{\bot} h_{ij}€}€
\end{equation}
where $g_{ij}({\bf k}_{\bot} )$ is  $g({\bf k}_{\bot}€) \equiv {2 J_{1} (k_{\bot} D) \over k_{\bot} D}$ 
if $i=j$ and 1 otherwise, and  $J_{1} (x)$  is 
the first-order Bessel function of the first kind.  Finally, $h_{ij}$ is the separation 
between plates $i$ and $j$

 First note that the free energy of an $N$-plate system is not simply 
a pairwise sum of the corresponding two-plate results over all pairs of 
plates.  Thus the pairwise additivity is not {\it always} satisfied as will 
be detailed later.  For the $N=2$ case, our result in Eq.~\ref{free}  
reduces to the previous two-plate result [see Eq.~3 of 
Ref.~\cite{pincus}], if $D$ is set to zero.   
When $D=0$, the free energy has a single minimum at a nonzero value  
$k_{\bot}=k_{\bot}^{<}  \ll 1 \AA^{-1}$ 
for all values of $\lambda$.  
 The dominance of the long-wavelength charge 
fluctuations is responsible for the breakdown of the pairwise additivity of 
electrostatic interactions between macroions.  It has been shown that 
the pairwise additivity for the case of charged rods  
breaks down if the expansion of the 
corresponding interaction free energy 
in powers of $\ell_{B}$ diverges~\cite{ha.2rod,ha.bundle,ha.pre}.  
For rod systems, the free energy is dominated by the zero-$k$ mode 
and thus this expansion converges {\it only when} the charge fluctuation 
along the rods is 
{\it vanishingly small}.  For the two-dimensional case, the 
convergence of the $\ell_{B}$-expansion  can 
be tested by estimating $\delta \equiv  (k_{\bot}^{<})^{-1}€ \lambda^{-1} $; 
the $\ell_{B}$-expansion is convergent if $\delta < 1$.  We, however, 
find that $\delta$ is smaller for smaller values of $\lambda^{-1}$ and 
is comparable to unity if $\lambda^{-1} < 10^{-3}$.  As in the 
rodlike  
systems, the pairwise additivity is violated unless the inverse of the 
Gouy-Chapman length is very small.  Thus the pairwise 
additivity can easily be violated even in two-dimensional systems.    
 \begin{figure}
   \par\columnwidth20.5pc
   \hsize\columnwidth\global\linewidth\columnwidth
   \displaywidth\columnwidth
   \epsfxsize=3.12truein
   \centerline{\epsfbox{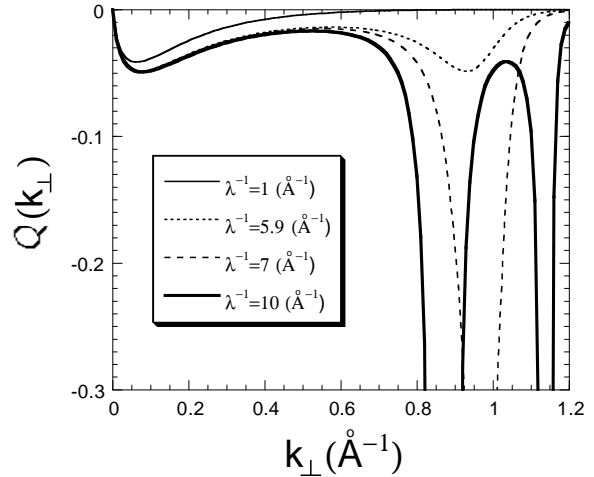}}
\caption{ The wavevector dependence of the interaction free energy 
  between two plates separated by $h=5 \AA$, as a function of 
  $k_{\bot}$.  
  We have chosen $D = 5 \AA$.  
  When $\lambda^{-1} = 1 \AA^{-1}€$, the free energy has 
 only one minimum at $ k_{\bot} \ll 1 \AA^{-1}€$,  
 while when $\lambda^{-1}€ =5.9 \AA^{-1}$ 
  the free energy has another local minima at $k_{\bot} = k_{\bot}^{>}€ 
  = {\cal O}(1 \AA^{-1}€)$.  
  When $\lambda^{-1}€=7 \AA^{-1}€$, the second minima at large $k_{\bot}$ is 
  overwhelmingly dominant.  When $\lambda^{-1}€=10 \AA^{-1}$  the 
  free energy has two local mimina 
  at  $k_{\bot} ={\cal O}(1 \AA^{-1})$.     
  \label{free(q).fig}}
  \end{figure}
  
The above analysis, however, cannot be pursued when the temperature 
is low or when the density of ions is high.  
In this case, it is crucial to incorporate 
$D \ne 0$~\cite{ha.correlation,ha.pre,ha.reply}; the ion size constitutes an 
important length scale at low 
temperatures or at high density of ions.  
Having understood that the 
 dominance of long-wavelength charge fluctuations for the case $D=0$  
  leads to the breakdown of pairwise additivity, we now 
 examine the low-$T$ behavior of the free energy  
 (thus with $D$ set to a finite value) in Fourier space.   
Let us first consider the case of two plates seperated by a distance $h$.        
For convenience, we consider the following quantity: 
${\cal Q}({\bf k}_{\bot}€)\equiv k_{\bot}  \log [  \det Q ({\bf 
k}_{\bot}€)]$, {\i.e.}, the first term in $\{\ldots\}$ of 
Eq.~\ref{free} mutiplied by 
$k_{\bot}$.   In 
Fig.~\ref{free(q).fig}, we plot this quantity ${\cal Q}({\bf k}_{\bot})$ 
as a function of $k_{\bot}$ for several 
different values of the Gouy-Chapman length $\lambda$.  We have 
chosen $D=5 \AA$ and $h=5 \AA$.  When 
$\lambda^{-1}€=1 \AA^{-1} $,  ${\cal Q}({\bf k}_{\bot}€) $ has a single minimum at 
$k_{\bot}=k_{\bot}^{<}€ \ll 1$.  This 
implies that the free energy is dominated by long-wavelength charge 
fluctuations as in the previous 
case of point charges.  We find that 
the function ${\cal Q}({\bf k}_{\bot}€)$ has two minima at 
$k_{\bot}= k_{\bot}^{<}
\ll 1 \AA^{-1}€ $ and at $k_{\bot} = k_{\bot}^{>}€ = {\cal O} (1 
\AA^{-1}€)$, respectively, for 
$\lambda^{-1} > 4.2 \AA^{-1}$ (not shown in the figure).  As $\lambda^{-1}$ changes, the 
minimum at  $k_{\bot}=k_{\bot}^{>}€$ varies monotonically and  is 
deeper for 
larger $\lambda^{-1}$.  The minimum at $k_{\bot}= k_{\bot}^{<}€$, however, 
is roughly independent of 
$\lambda$.  When $\lambda^{-1} \approx 5.9 \AA^{-1}€$, the two minima are 
comparable in magnitude.  When $\lambda^{-1}= 7 \AA^{-1}€$, the function 
${\cal Q}({\bf k}_{\bot}€)$ is overwhelmingly dominated by the second 
minimum at $k_{\bot}= k_{\bot}^{>}$, as shown in the figure.  At
 $\lambda^{-1} = \lambda^{-1}_{X}=7.2 \AA^{-1}$, the second minimum 
 diverges.  This is suggestive of the onset of crystallization of 
 counterions as will be detailed later.      Even though the 
 region $\lambda^{-1} \geq \lambda_{X}^{-1}$ 
is certainly beyond the validity of our theory, it is neverthless 
interesting to see what our theory impiles for that region.   
 Notably, the free energy curve 
corresponding to $\lambda^{-1}=10 \AA^{-1}$ has two 
local minima at large $k_{\bot}={\cal O} (1 \AA^{-1}€)$. 
The existence of multiple minima at large $k_{\bot}$ 
assures that the system is in a solidlike phase.

Our results in Fig.~\ref{free(q).fig} imply that there are two distinct 
contributions to the free energy: long-wavelength fluctuations and 
short-wavelength fluctuations in the density of counterions.  They 
also imply that the short-wavelength fluctuation contribution to the 
free energy has a much narrower peak if $\lambda^{-1} \gg 5.9 
\AA^{-1}€$.  This enables us to separate the 
short-wavelength contribution from the long-wavelength contribution.  
By noting that ${\rm e}^{-k_{\bot} h}$ does not 
change appreciably over the region inside the peak at $k_{\bot}€=k_{\bot}^{>}€$, we 
find, up to $h$-independent terms  
\begin{eqnarray}
\label{free.asymp}
&&\Delta F_{2}€ \simeq  -{k_{B}€T \over 16 \pi} {\zeta (3) \over 
h^{2}€} \nonumber \\
&&\quad - 
{ k_{B}T  \over 8 \pi^{2}€} {{\rm e}^{-2 k_{\bot}^{>}€ h } \over 
\lambda^{2}}€ 
\int_{k_{\bot} \simeq  k_{\bot}^{>}€ } 
 {k_{\bot}^{-1}€d k_{\bot}
 \over [1+k_{\bot}^{-1}€ \lambda^{-1}€g({
k}_{\bot})]^{2}€ } 
,\end{eqnarray}
where $\zeta (x)$ is the zeta function (thus $\zeta (3)/16 \pi \simeq 
0.024$).  The first term denoted by $ F_{LW} $ 
is the free energy calculated with $D$ set to 
zero~\cite{kardar,pincus,safran,attard} and is  the long-wavelength 
free energy.  Our previous analysis on ${\cal Q}({\bf 
k}_{\bot}€)$ implies that the short-wavelength  free 
energy denoted by $ F_{SW}$, {\i.e.}, the second term 
in Eq.~\ref{free.asymp}, is dominant over the long-wavelength free energy 
$ F_{LW}$ at low temperatures  
($\lambda^{-1} \gg 5.9 \AA^{-1}$) and decays 
exponentially in space.  The exponentially decaying interaction between two 
plates is consistent with the zero-$T$ analysis in 
Refs.~\cite{bloomfield,lau.plasmon}.    At high temperatures  
corresponding to $\lambda^{-1} \ll 5.9 \AA^{-1}$, however, the free energy is 
mainly determined by $F_{LW}€$.  In this case, the fluctuation 
contribution to the pressure between the two plates scales as $- 
h^{-3}$~\cite{kardar,pincus,safran,attard}.  Also note that this long-wavelength contribution 
vanishes at $T=0$.  This follows from 
the fact that ${\cal Q}(k_{\bot}^{<}€)$ is roughly independent of 
$\lambda$.  To estimate the temperature dependence of $F_{SW}€$, note that the 
prefactor of this term varies as 
$T^{-1}€$.  The $k_{\bot}$-integral of this term depends 
on the depth and width of the second minimum of ${\cal Q}(k_{\bot})$ at 
$k_{\bot}=k_{\bot}^{>}€$.  While the width is roughly  independent of 
$\lambda$ for given $h$, the minimum becomes deeper with the 
increasing $\lambda^{-1}$ (or decreasing $T$).    This proves that  $ 
F_{SW}$ is more negative at low $T$ than at high $T$, as opposed to $ 
F_{LW}$. 

\begin{figure}
   \par\columnwidth20.5pc
   \hsize\columnwidth\global\linewidth\columnwidth
   \displaywidth\columnwidth
   \epsfxsize=3.12truein
   \centerline{\epsfbox{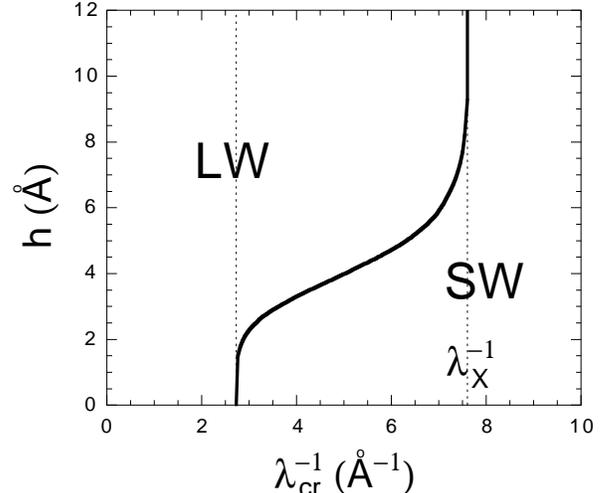}} 
\caption{Phase diagram for two plates.  The regimes where long-wavelength and 
short-wavelength fluctuations dominate are denoted by LW and SW, respectively.  
\label{crossover}}
\end{figure}

For a given value of $h$, there exists a special value $\lambda_{cr}$ 
at which the crossover between the two distinctive behaviors of the 
plate interaction ($F_{LW}$ 
and $F_{SW}$) takes place.   By requiring ${\partial  \over \partial h}
\left(F_{LW}-F_{SW}€\right)=0$, we have the following transcendental equation 
for $\lambda_{cr}$:
\begin{equation}
\left({\pi \over 2} \right) {\zeta (3) \over 
k_{\bot}^{>}} {{\rm e}^{2 k_{\bot}^{>} h} 
\over h^{3}€} 
={1€ \over \lambda^{2}_{cr}€}  \int_{k_{\bot} \simeq  k_{\bot}^{>}€ } 
 {k_{\bot}^{-1}€d k_{\bot}
 \over [1+k_{\bot}^{-1}€ \lambda^{-1}_{cr} g({
k}_{\bot})]^{2}€ } 
.\end{equation}  
To solve the trascendental equation, we have chosen  $D=5 \AA$.    
Fig.~\ref{crossover} describes distinct regimes characterized by the 
corresponding dominant modes of fluctuations, and the crossover boundaries 
between them; the regimes where 
the long-wavelength (LW) and short-wavelength (SW) fluctuations dominate 
are denoted by LW and SW, respectively.   When $\lambda^{-1}€$ is smaller 
than $2.7 \AA^{-1}€$,  the plate interaction is solely determined by the 
LW fluctuations for the whole range of $h$.   
At $\lambda^{-1} \simeq 2.7 \AA^{-1}$, marked by the vertical dotted line 
on the left, the SW fluctuations start to contribute to the plate interaction.  
When  $2.7 \AA^{-1} \le \lambda^{-1} \le 7.6 \AA^{-1}$, however, the 
plate interaction is determined by the competition between the two; the crossover 
from the LW to SW regime takes places for larger value of $h$ at low 
temperatures (corresponding to larger $\lambda^{-1}$).  At 
$\lambda^{-1} =\lambda^{-1}_{X}€\simeq 7.6 \AA^{-1}€$~\cite{remark}, 
marked by the  vertical dotted line on the right, the crossover  occurs 
only when $h \rightarrow \infty $.     
 Beyond $\lambda_{X}^{-1} $,  the SW fluctuations solely determine the plate 
 interaction.    
The appearance of two distinctive competing interactions, {\i.e.}, $ 
F_{LW}$ and $F_{SW}€$, can also be understood in terms of  in-plane charge 
correlations for a single plate: $G_{1 }€({\bf r}_{\bot}€,{\bf 
r}_{\bot}^{\prime}) =   
\left< \hat \sigma ({\bf r}_{\bot}) \hat \sigma ({\bf r}_{\bot}^{\prime}) \right>- 
 \left< \hat \sigma ({\bf r}_{\bot}) \right> \left< \hat \sigma ({\bf 
 r}_{\bot}^{\prime}) 
 \right>$.  
The long-wavelength contribution to the charge correlation was shown 
to scale as $G_{LW} ({\bf r}_{\bot},{\bf r}_{\bot}^{\prime} ) \sim 
-k_{B}T  /|{\bf r}_{\bot}-{\bf r}_{\bot}^{\prime} 
|^{3}$~\cite{attard}.   
Note that this 
correlation vanishes as $T \rightarrow 0$.  Our result in 
Eq.~\ref{free} implies that the short-wavelength charge correlation function  is 
given by
\begin{eqnarray}
\label{G(R)}
&&G_{SW}({\bf r}_{\bot},{\bf r}_{\bot}^{\prime})  
 \simeq g({\bf r}_{\bot}€,{\bf r}_{\bot}^{\prime}€)  \nonumber \\
 &&\quad - {\sigma_{0} e^{2} \over  2  \pi }
  \int_{k_{\bot}€\ne 0} 
  {  k_{\bot} d  k_{\bot}  g ({\bf k}_{\bot}) \over 
    1+{\lambda { k}_{\bot} \over g({\bf k}_{\bot}) } }
    J_{0}(k_{\bot} | {\bf r}_{\bot}-{\bf r}_{\bot}'| )
    ,\end{eqnarray}
where $J_{0}(x)$ is the zeroth-order  Bessel function of the first 
kind.  Unlike the long-wavelength 
correlation, the behavior of the short-wavelength correlation is determined by the 
nature of the poles of $\left[1+\lambda k_{\bot }/g({\bf k}_{\bot}€) 
\right]^{-1}$.  We find that $G_{SW}€$ shows an oscillatory decay, and 
that the amplitude of $G_{SW}$ varies as $T^{-1} \left[ 1+ 
(k_{\bot}^{>} \lambda)^{-1}€ g(k^{>}_{\bot})  \right]^{-1}€$ and is 
larger at low temperatures.  For the two-plate case, this oscillatory 
charge correlation essentially gives rise to the SW interaction.

At high temperatures,  we can 
consider each 
plate to consist of large domains, {\i.e.}, counterion-rich and 
counterion-poor domains.   The size of the domains is on the order of 
$(k_{\bot}^{<})^{-1}€ \gg 1 \AA$, and thus these domains can form huge dipoles, 
resulting in a long-ranged attraction.  The long-wavelength fluctuations couple over 
many  plates, leading to breakdown of pairwise 
 additivity~\cite{remark.correlation} in the $N$-plate case.  We also find that the 
 power-law behavior of 
 the in-plane charge correlation crosses over to  
  an exponential decaying form as 
 $N \rightarrow \infty$~\cite{unpublished}.  This is clearly 
 suggestive of the breakdown of pairwise additivity.  
 As the temperature decreases, however, 
 the plate interaction driven by long-wavelength fluctuations crosses 
 over to a distinct behaviour controlled by non-monotonically decaying charge 
 correlations.  At low temperatures,  each 
 domain becomes overall charge neutral and thus the distiction between domains is 
 meaningless.  In this case, the local correlation between a 
 counterion and a backbone charge in its neighborhood dominates  the 
 free energy.  There is thus strong 
 cancellation of  repulsions (between like charges) with attractions 
 (between opposite charges).  This results in an exponentially decaying, short-ranged 
 attraction between the plates.

In conclusion, we have presented a systematic approach to study the
 nature of counterion-mediated attractions between fluid membranes.  We have
 shown that the nature of these attractions is determined by the
 dominant modes of fluctuations in the density of counterions.   At high
 temperatures, fluctuations at large length scales determine the membrane
 interactions; the resulting interactions are long-ranged and not
 pairwise additive.  Charge densities of biomembranes range from $0.03$ to 
 $0.24/{\rm nm}^{2}$, corresponding to LW regimes 
 at room temperature.  It is thus clear that  many-body, non-pairwise additive 
 interactions operate in biomembrane systems at room temperature.  
 At low temperatures, however, the
 membrane interactions are dominated by SW charge
 fluctuations  {\it and}
 are {\it exponentially} decaying with the membrane spacing.  In this case, only
 the nearest pairs of plates couple strongly with each other.  Surprisingly, this
 implies that the pairwise additivity is restored at very low 
 temperatures (the
 non-pairwise additive interaction becomes smaller, and eventually
vanishes, as $T \rightarrow 0$).   The approach presented here allows one
 to systematically study the crossover, as the temperature decreases, from the
high-temperature,  long-ranged attractions to the behaviors expected at low 
 temperatures.     

We have benifited from illuminating discussions with A.W.C. Lau and A.J. 
Liu.  We are also grateful to W.M. Gelbart, H. Schiessel, and R. Menes 
for scientific stimulation.  We thank M. Howard for reading the 
manuscript carefully, and  
D. Boal, M. Wortis, and M. Plischke for providing an excellent research enviornment.
  This work was in part supported by the NSERC of Canada.

\end{multicols}


\begin{references}
\bibitem{remark} Note that this value is somewhat larger than the 
corresponding value for two plates, {\i.e.}, $\lambda^{-1}_{X} \simeq 
7.2 \AA^{-1}$.
\end{references}
\end{document}